\documentclass[
]{ceurart}

\usepackage[T1]{fontenc}
\usepackage{lmodern}

\begin{document}

\copyrightyear{2022}
\copyrightclause{Copyright for this paper by its authors.
  Use permitted under Creative Commons License Attribution 4.0
  International (CC BY 4.0).}

\conference{Proceedings of the First Workshop on Automatic Speech Recognition for Spontaneous and Prepared Speech and Speech emotion recognition in Portuguese (SER 2022), co-located with PROPOR 2022.  March 21st, 2022 (Online).}

\title{Pretrained audio neural networks for Speech emotion recognition in Portuguese}

\author[1]{Marcelo Matheus Gauy}[%
email=marcelomatheusgauy@gmail.com,
]
\address[1]{Universidade de S\~{a}o Paulo,
  Rua do Mat\~{a}o 1010, S\~{a}o Paulo, Brazil}

\author[1]{Marcelo Finger}[%
email=mfinger@ime.usp.br,
]

\begin{abstract}
  The goal of speech emotion recognition (SER) is to identify the emotional aspects of speech. The SER challenge for Brazilian Portuguese speech was proposed with short snippets of Portuguese which  are classified as neutral, non-neutral female and non-neutral male according to paralinguistic elements (laughing, crying, etc). This dataset contains about $50$ minutes of Brazilian Portuguese speech. As the dataset leans on the small side, we investigate whether a combination of transfer learning and data augmentation techniques can produce positive results. Thus, by combining a data augmentation technique called SpecAugment, with the use of Pretrained Audio Neural Networks (PANNs) for transfer learning we are able to obtain interesting results. The PANNs (CNN6, CNN10 and CNN14) are pretrained on a large dataset called AudioSet containing more than $5000$ hours of audio. They were finetuned on the SER dataset and the best performing model (CNN10) on the validation set was submitted to the challenge, achieving an $F1$ score of $0.73$ up from $0.54$ from the baselines provided by the challenge. Moreover, we also tested the use of Transformer neural architecture, pretrained on about $600$ hours of Brazilian Portuguese audio data. Transformers, as well as more complex models of PANNs (CNN14), fail to generalize to the test set in the SER dataset and do not beat the baseline. Considering the limitation of the dataset sizes, currently the best approach for SER is using PANNs (specifically, CNN6 and CNN10).
\end{abstract}

\begin{keywords}
  Speech emotion recognition \sep
  Pretrained audio neural networks \sep
  Transfer learning \sep
  Transformers
\end{keywords}

\maketitle

\section{Introduction}

Speech emotion recognition (SER) aims at identifying the emotional aspects of speech independently from the actual semantic content. SER can be used to identify the emotions of humans, e.g.,  when using mobile phones, an ability that may become crucial in improving human-machine interactions in the future~\cite{andre2004endowing}. Several efforts to acquire speech data classified with different emotional labels have been undertaken~\cite{livingstone2018ryerson, wang2010machine, busso2008iemocap}. These datasets are typically small in size, even for languages such as English. In order to tackle these datasets, the use of transfer learning and data augmentation techniques may be instrumental.

Transfer learning is the method of training a network on a particular problem where there is an abundance of data, with the goal of using the acquired knowledge to obtain better performance on a related problem with limited data available. Transfer learning has been effectively used in many fields of deep learning such as computer vision~\cite{voulodimos2018deep} and language modelling~\cite{devlin2018bert}. Data augmentation is the method of increasing the amount of data available by slightly modifying copies of the data. This can be done, for example, by masking parts of the input or by adding Gaussian noise to it.

In this paper, we use transfer learning and data augmentation techniques to study SER in Brazilian Portuguese speech. We participate in the \textit{shared task SER challenge}, a challenge for Brazilian Portuguese speech emotion recognition. This challenge made available a labeled dataset of $625$ audio files as training set for SER. Moreover, a dataset of $308$ files was made available as the test set. The training and test datasets consisted of short  snippets of Brazilian Portuguese speech, usually less than $15\,s$ long, labeled neutral, non-neutral female and non-neutral male (non-neutral for audios containing laughs, cries, etc).

For transfer learning, we employ \textit{Pretrained audio neural network (PANN)}~\cite{kong2020panns}, which are convolutional neural networks trained on a large dataset of audios (AudioSet~\cite{gemmeke2017audio}), consisting of $1.9$ million audio clips distributed across $527$ sound classes. By using the pretrained models made available by the developers, and finetuning on the SER dataset for Brazilian Portuguese speech, we are able to beat the \href{https://sites.google.com/view/ser2022/shared-tasks}{proposed baselines} of prosodic features and wav2vec features. We achieve (via CNN10) F1-score of $0.73$, up from $0.54$ from the baselines. During finetuning, we employ a data augmentation technique called SpecAugment~\cite{park2019specaugment}.

We also tested the use of Transformer neural networks, pretrained on a large amount of Brazilian Portuguese audio data~\cite{gauy2022acoustic}. However, we find that, with the current amount of available data for SER, Transformers do not generalize their training performance to the validation and test sets. This holds even while using most common techniques to prevent overfitting. The same behaviour was also observed for more complex PANNs, such as CNN14.

\section{Related Work}

There is a large literature on SER in English~\cite{lech2020real, yoon2019speech, xu2019learning, yoon2018multimodal, satt2017efficient, issa2020speech, peng2021efficient, pepino2021emotion}. Moreover, there are a lot of small datasets for SER in English, such as, RAVDESS~\cite{livingstone2018ryerson}, SAVEE~\cite{wang2010machine} and IEMOCAP~\cite{busso2008iemocap}. To the best of our knowledge, the SER dataset for Brazilian Portuguese speech is the only available dataset on the language. In addition, English datasets are usually classified in a different set of labels. RAVDESS~\cite{livingstone2018ryerson}, for example, has the classes of calm, happy, angry, sad, fearful, surprise and disgust. This contrasts with the classes of neutral, non-neutral female and non-neutral male present in the SER dataset for Brazilian Portuguese speech. As such, comparisons of our work with the state of the art in English language are not really possible. Nevertheless, the authors of~\cite{pepino2021emotion}, the most recent work, obtain an average recall on RAVDESS of $84.3$ percent using wav2vec 2.0~\cite{baevski2020wav2vec}. On IEMOCAP, they obtain an average recall of $67.2$ percent, also using wav2vec 2.0.

Transfer learning is a very common technique in situations where the dataset available is small in size. It has been effectively employed in computer vision~\cite{voulodimos2018deep, krizhevsky2012imagenet}, language modelling~\cite{devlin2018bert, brown2020language} and audio tasks~\cite{kong2020panns, wang2015transfer, pepino2021emotion}. In the original PANN paper~\cite{kong2020panns}, authors propose several convolutional neural networks pretrained on AudioSet which can be finetuned on other smaller datasets. In~\cite{pepino2021emotion} the authors use wav2vec 2.0 pretrained on Librispeech and finetuned on either RAVDESS or IEMOCAP for speech emotion recognition. Finally, in~\cite{wang2015transfer} the authors provide a comprehensive review on transfer learning methods used for speech and language processing tasks.

\section{Methodology}

The code for this paper can be found at~\href{https://github.com/marcelomatheusgauy/Pretrained_audio_neural_networks_emotion_recognition}{GitHub}. Below we describe the dataset and architectures used.

\subsection{SER Dataset}

To perform SER on Brazilian Portuguese speech, we use the training dataset (CORAA SER version $1.0$) provided for the challenge. This dataset was built from the C-ORAL-BRASIL I corpus~\cite{raso2012c}, with $625$ audio files, typically less than $15s$-long, containing informal spontaneous Brazilian Portuguese speech. These audio files are labeled neutral, non-neutral female, non-neutral male. An audio is labeled non-neutral male if it is a male speaker and it contains paralinguistic elements in the speech (such as laughing, crying, etc). Similarly, an audio is labeled non-neutral female if it is a female speaker and the speech contains such paralinguistic elements.

We split the official training dataset into training ($80\%$), validation ($10\%$) and test sets ($10\%$). The split was done in an arbitrary way to ensure that the three datasets were balanced (i.e. contained relatively the same proportion of neutral, non-neutral female and non-neutral male files). The training dataset consisted of $500$ files, the validation dataset consisted of $63$ files and the test set of $62$ files. The results we report are for the validation and test set performance.

As the official test dataset made available did not have labels,  we have labeled it ourselves, out of curiosity and to enable more consistent tests of the performance of the networks. While the labels may not be perfect, they provide a close enough picture, so the performance of the models can be measured as an average over multiple experiments (as we were observing high variance). As such, we also provide results for the official test set with our unofficial labels. We stress that we did not use the test set labels for any form of model or parameter selection.

Lastly, the PANNs we use have been trained on the AudioSet~\cite{gemmeke2017audio} dataset containing more than $5000$ hours of audio distributed across $527$ classes. 

\subsection{PANN Architectures}

Table~\ref{table:architectures} describes the three architectures we use. They are named CNN6, CNN10 and CNN14 after the $6$-layer, $10$-layer and $14$-layer CNNs they represent. These are the same CNN network architectures used in~\cite{kong2020panns}. We take their pretrained models on AudioSet~\cite{gemmeke2017audio} to allow us to obtain better generalization performances on the SER dataset.

The audios are preprocessed in the following way: the audios are first resampled to $32kHz$. After that, we apply short-time Fourier transform~\cite{brigham1967fast} (with a window size of $1024$ frames and hop size of $320$ frames) to the standard time-domain waveforms to obtain spectrograms. Then, Mel filter banks are applied to spectrograms, followed by a logarithm operation to obtain log Mel spectrograms. These preprocessing steps are commonly done when using CNNs for audio~\cite{choi2016automatic, kong2019weakly}.

As described in Table~\ref{table:architectures}, the CNN architectures used are composed of convolutional layers with kernel $5\times 5$ for CNN6, and $3\times 3$ for CNN10 and CNN14. Each convolutional layer is followed by batch normalization~\cite{ioffe2015batch} and ReLU non-linearity~\cite{nair2010rectified} is used to allow for better training convergence. Each such convolutional block is present $4$ times in CNN6 and, in between, an average pooling $2\times 2$ layer is applied (average pooling is observed to be better than max pooling~\cite{kong2019cross}). In CNN10 and CNN14, the convolutional blocks are always used in pairs before an average pooling layer is applied. CNN10 contains $8$ such convolutional blocks ($4$ pairs) and CNN14 contains $12$ such convolutional blocks ($6$ pairs). All networks have a penultimate fully connected layer to add extra representation ability, as well as a final $527$ units fully connected layer where a sigmoid is applied to obtain the probabilities for each class. In Table~\ref{table:architectures}, the first line describes the input of the networks, that is, $n$ frames of Log Mel Spectrogram with $64$ mel bins for each frame. Each subsequent line represents a layer of the networks. The numbers following the @ sign represent the quantity of $5\times 5$ or $3\times 3$ feature maps used.
 
\begin{table}
\caption{PANN architectures. We describe the layers of CNN6, CNN10 and CNN14.} \label{table:architectures}
\begin{tabular}{|m{7em}|m{7em}|m{7em}|}
\hline
CNN6 & CNN10 & CNN14 \\ \hline
\multicolumn{3}{|c|}{Log Mel Spectrogram $n$ frames $\times$ 64 mel bins} \\ \hline
$\binom{ 5 \times 5 @ 64} {BN, ReLU}$ & $\binom{ 3 \times 3 @ 64} {BN, ReLU} \times 2$ & $\binom{ 3 \times 3 @ 64} {BN, ReLU} \times 2$ \\ \hline
Avg Pooling $2\times 2$ & Avg Pooling $2\times 2$ & Avg Pooling $2\times 2$ \\ \hline
$\binom{ 5 \times 5 @ 128} {BN, ReLU}$ & $\binom{ 3 \times 3 @ 128} {BN, ReLU} \times 2$ & $\binom{ 3 \times 3 @ 128} {BN, ReLU} \times 2$ \\ \hline
Avg Pooling $2\times 2$ & Avg Pooling $2\times 2$ & Avg Pooling $2\times 2$ \\ \hline
$\binom{ 5 \times 5 @ 256} {BN, ReLU}$ & $\binom{ 3 \times 3 @ 256} {BN, ReLU} \times 2$ & $\binom{ 3 \times 3 @ 256} {BN, ReLU} \times 2$ \\ \hline
Avg Pooling $2\times 2$ & Avg Pooling $2\times 2$ & Avg Pooling $2\times 2$ \\ \hline
$\binom{ 5 \times 5 @ 512} {BN, ReLU}$ & $\binom{ 3 \times 3 @ 512} {BN, ReLU} \times 2$ & $\binom{ 3 \times 3 @ 512} {BN, ReLU} \times 2$ \\ \hline
Global Avg Pooling & Global Avg Pooling & Avg Pooling $2\times 2$\\ \hline
FC $512$, ReLU & FC $512$, ReLU & $\binom{ 3 \times 3 @ 1024} {BN, ReLU} \times 2$\\ \hline
FC $527$, Sigmoid & FC $527$, Sigmoid & Avg Pooling $2\times 2$\\ \hline
\multicolumn{2}{|c|}{} & $\binom{ 3 \times 3 @ 2048} {BN, ReLU} \times 2$ \\ \cline{3-3}
\multicolumn{2}{|c|}{} & Global Avg Pooling \\ \cline{3-3}
\multicolumn{2}{|c|}{} & FC $2048$, ReLU \\ \cline{3-3}
\multicolumn{2}{|c|}{} & FC $527$, Sigmoid \\ \hline
\end{tabular}
\end{table}

\subsection{Transformer Encoder Architecture}

In addition to experimenting with the PANNs, we also attempt to extract good performances from Transformers. The Transformer architecture we use is equivalent to the Transformer Encoder architecture from~\cite{vaswani2017attention}. That is, we use a three-layer Transformer with multi-head self attention. Each encoder layer is composed of two sub-layers. The first is a multi-head self-attention network and the second is a fully connected feed-forward layer. Each sub-layer has a residual connection followed by layer normalization~\cite{ba2016layer}. The encoder layers and sub-layers produce outputs of dimension $d$ (in experiments $d$ is either $128$ or $512$). The fully connected feed forward network within each encoder layer has an inner dimension of $4d$. We feed the Transformer Encoders the MFCC-gram of the audios, with each token fed to the Transformer corresponding to a frame of the MFCC-gram~\cite{219270}. We name these Transformers, the MFCC-gram Transformers~\cite{219270}. We use sinusoidal positional encoding so the Transformer has access to the order of the sequence fed~\cite{vaswani2017attention, liu2020mockingjay}. The input frames are projected linearly to a hidden layer of dimension $d$, as direct addition of acoustic features to positional encoding may lead to training failure~\cite{liu2020mockingjay}.

Typically, Transformers undergo two training phases: pretraining and finetuning. In the pretraining phase, we make use of a technique called Time alteration~\cite{liu2020mockingjay} to pretrain the Transformer in about $600$ hours of Brazilian Portuguese audio data (in other words, we use pretrained models from~\cite{gauy2022acoustic}). Time alteration is a technique that masks random spans of frames of the MFCC-gram similarly to how time masking functions in SpecAugment (described in subsection~\ref{subsec:specaugment}). During pretraining, the model is trained to reconstruct the masked frames. For Brazilian Portuguese audio data, we use the corpora of NURC-S\~{a}o Paulo\cite{castilho1986nurc}, NURC-Recife~\cite{oliviera2016nurc}, ALIP~\cite{gonccalves2019projeto}, SP2010~\cite{mendes2013projeto} and Programa Certas Palavras~\cite{teixeira1997certaspalavras}. In the experiments, we also show the performance of Transformers which do not undergo pretraining, that is, which we initialize at random and do finetuning directly. We name those Transformers the Baseline MFCC-gram Transformers. After pretraining, the Transformers are finetuned on the SER dataset.

\subsection{Data augmentation: SpecAugment}\label{subsec:specaugment}

The SER training dataset used for the challenge leans on the small side and contains about $50$ minutes of audio. To mitigate the potential overfitting effects of a small training dataset, we perform a common audio data augmentation technique called SpecAugment~\cite{park2019specaugment} on the Mel spectrogram (or MFCC-gram) of the audio files before feeding it to the network's layers. SpecAugment consists in masking random spans of consecutive segments of the spectrogram of the audios. Masking can be done along the time dimension (that is, on spans of consecutive frames), or along the frequency dimension (that is, on spans of consecutive frequency channels).

Following~\cite{kong2020panns}, time masking is done by selecting a uniform length $\ell$ (chosen between $0$ and $64$) and a uniform frame start $t$ (chosen between $0$ and $T-\ell$, where $T$ is the total number of frames of the audio) and proceeding to mask the frames from $t$ to $t+\ell-1$. We mask two such blocks of consecutive frames. Frequency masking is similar to time masking but done along the frequency dimension. So, a random uniform length $\ell$ is chosen (between $0$ and $8$) and a uniform frequency band $f$ is chosen (between $0$ and $F-\ell$ where $F$ is the total number of Mel frequency bins). The frequency bands from $f$ to $f+\ell-1$ are masked to zero. As with time masking, we mask two such blocks of consecutive frequency bands.

\section{Results and Discussion}

We will check the performance of the three proposed PANNs (CNN6, CNN10 and CNN14) on the SER training and test datasets. In order to take advantage of the large pretraining done on the AudioSet~\cite{gemmeke2017audio} dataset, we will use the pretrained models of CNN6, CNN10 and CNN14 made available by the authors of~\cite{kong2020panns}. These can be found in~\href{https://zenodo.org/record/3987831}{Zenodo}. These pretrained models will be finetuned on the SER training dataset in order to achieve better performance than the baseline. 

Moreover, to showcase the massive level of transfer learning that is happening via the pretrained models, we will show the performance of the three networks (CNN6, CNN10 and CNN14) without the use of a pretrained model, that is, initializing their weights at random and not making use of the AudioSet~\cite{gemmeke2017audio} pretraining. We call these three models the Baseline CNN6, the Baseline CNN10 and the Baseline CNN14.

Lastly, we show the performance of three Transformers models. We analyze MFCC-gram Transformers pretrained on about $600$ hours of Brazilian Portuguese audio data, as well as, Baseline MFCC-gram Transformers (without pretraining) containing $512$ and $128$ units per Encoder layer.

As mentioned before, the SER training dataset is split into a training ($80\%$), validation ($10\%$) and test sets ($10\%$). In Table~\ref{table:results}, we report the $F1$ score performance of the nine models in the validation and test datasets as well as in the official dataset (which was labeled by us). The results in the table are averaged across $25$ experiments, to better control the generally high $F1$ score variance between different experiments. Each experiment consisted of training the model for $100$ epochs for CNNs and $20$ epochs for Transformers\footnote{As Transformer does not generalize, no advantage exists in training it for longer than $20$ epochs.} in the training set and the best validation performance model (checked after each epoch) was saved and later analyzed on the test set and official test set. The batch size used was $16$ and the learning rate was $10^{-4}$ for the CNNs and we use a warmup learning rate schedule according to the formula $d^{-0.5}\times min(stepnumber^{-0.5}, stepnumber\times warmupsteps^{-1.5})$ for the Transformers as is standard~\cite{devlin2018bert}. We use $warmupsteps=4000$.

As can be seen on Table~\ref{table:results}, the best results in the test set were attained by the CNN6 ($0.62$ $F1$ score). Moreover, it seems that the test set built by us was inherently harder than the official test set. In the official test set, the best result was obtained by CNN10 ($0.74$ $F1$ score), in line with it achieving also the best results on the validation set.

We observe that CNN14's performance was significantly worse both on validation and test. However, representation ability wise it is the most powerful of the PANN models. It is likely that the SER dataset being so small meant CNN14 suffered from overfitting. 

We also experienced overfitting issues when attempting MFCC-gram Transformers based models. There, using pretraining techniques did not yield better performance. This is likely because the pretraining data contained primarily voice, without laughs or cries, so the important markers were not present in pretrained data. Moreover, no common technique (such as dropout~\cite{srivastava2014dropout}, L1 or L2 regularization~\cite{goodfellow2016deep}, data augmentation techniques as SpecAugment~\cite{park2019specaugment} and Mixup~\cite{zhang2017mixup}) to prevent overfitting yielded good results. It seems that the reduced size of the SER dataset is currently hindering performance in more complex networks, so a likely way of dramatically improving results would be to increase the size of the available dataset.

Lastly, note that the three baseline PANN models are far away from beating the baselines provided by the challenge. There is noticeable transfer learning benefit in using the pretrained models on AudioSet~\cite{gemmeke2017audio}. This large difference illustrates again the fact that the SER dataset is so small ($50$ minutes of audio) and that these networks suffer to generalize on it.

We have sent for evaluation in the challenge, the model which attained best test performance (a CNN6 which officially reported $0.66$ $F1$-score) and the model which attained best validation performance (a CNN10 which officially reported $0.73$ $F1$-score). Moreover, out of curiosity, we show the confusion matrix of the CNN10 model sent for evaluation in Table~\ref{table:confusion_matrix}. Observe that the model classifies the vast majority of neutral and non-neutral females files correctly. Most of the errors are done classifying non-neutral male files (often wrongly classified as neutral).


\begin{table}
\caption{The mean and standard deviation of the $F1$ score is shown in the table below for the nine models (CNN6, CNN10 and CNN14 and their respective baseline version, i.e., their versions without pretraining on AudioSet~\cite{gemmeke2017audio}, as well as MFCC-gram Transformers with and without pretraining and a smaller version of MFCC-gram Transformers). The results shown are for the validation set, the test set and the official test set. Labels for the official test set were created by us.}\label{table:results}
\begin{tabular}{|m{6em}|m{10em}|m{10em}|m{10em}|}
\hline
Model & $F1$ score Validation performance & $F1$ score Test performance & $F1$ score Official test performance \\ \hline

Baseline CNN6  & $0.45\pm 0.06$           & $0.36\pm 0.05$ & $0.33\pm 0.03$ \\ \hline
Baseline CNN10 & $0.58\pm 0.06$  & $0.41\pm 0.09$          & $0.42\pm 0.05$ \\ \hline
Baseline CNN14 &  $0.38\pm 0.06$          & $0.33\pm 0.04$          & $0.32\pm 0.03$ \\ \hline
CNN6  & $0.78\pm 0.05$           & $\mathbf{0.62\pm 0.06}$ & $0.69\pm 0.04$ \\ \hline
CNN10 & $\mathbf{0.80\pm 0.06}$  & $0.57\pm 0.06$          & $\mathbf{0.74\pm 0.04}$ \\ \hline
CNN14 &  $0.61\pm 0.11$          & $0.54\pm 0.06$          & $0.52\pm 0.10$ \\ \hline
MFCC-gram Transformers $512$ units &  $0.50\pm 0.04$          & $0.36\pm 0.06$          & $0.38\pm 0.03$ \\ \hline
Baseline MFCC-gram Transformers $512$ units &  $0.57\pm 0.04$          & $0.43\pm 0.08$          & $0.43\pm 0.06$ \\ \hline
Baseline MFCC-gram Transformers $128$ units &  $0.60\pm 0.05$          & $0.45\pm 0.07$          & $0.44\pm 0.04$ \\ \hline

\end{tabular}
\end{table}

\begin{table}
\caption{We plot the confusion matrix for the CNN10 model which was submitted to the challenge and attained an $F1$ score of $0.73$ (on official labels). Note that the model has the most difficulty classifying non-neutral male files correctly.}\label{table:confusion_matrix}
\begin{tabular}{|m{6em}|m{10em}|m{10em}|m{10em}|}
\hline
Confusion Matrix & predicted neutral & predicted non-neutral male & predicted non-neutral female \\ \hline

Neutral  & $244$           & $2$ & $5$ \\ \hline
Non-neutral male & $14$  & $8$          & $2$ \\ \hline
Non-neutral female &  $6$          & $1$          & $26$ \\ \hline

\end{tabular}
\end{table}

\section{Conclusion}

In this paper, we have effectively used transfer learning to beat the proposed baselines in the shared task SER challenge in Brazilian Portuguese speech. By using, the PANNs CNN6 and CNN10, we have attained $F1$ score of $0.73$ up from $0.54$ from the baselines. We have also observed that more complex networks, such as CNN14 and Transformers, while being in theory more capable of attaining better performances, suffer from overfitting. As such, we determine that probably the best way of improving results is by increasing the size of the training set.

Future work could involve increasing the size of the training set so that Transformers and CNN14 generalize their training performances to the test set. In addition, pretraining Transformers with audio data containing specifically laughs, cries and so on may prove useful. Moreover, other data augmentation techniques could be used which might provide additional benefit in terms of preventing overfitting.

\begin{acknowledgments}
This work was supported by FAPESP grant number 2020/16543-7 (POSDOC) and project 06443-5 (SPIRA).  This work was carried out at the Center for Artificial Intelligence (C4AI-USP), with support by the São Paulo Research Foundation (FAPESP) (grant \#2019/07665-4) and by the IBM Corporation. Marcelo Finger was partly supported by the São Paulo Research Foundation (FAPESP) (grants \#2015/21880-4, \#2014/12236-1); and the National Council for Scientific and Technological Development (CNPq) (grant PQ 303609/2018-4).  This work was financed in part by the Coordenação de Aperfeiçoamento de Pessoal de Nível Superior -- Brasil (CAPES) -- Finance Code 001.
\end{acknowledgments}

\bibliography{sample-ceur}

\appendix

\end{document}